\documentclass[9pt,twocolumn,twoside]{pnas-new}

\usepackage{graphicx}
\usepackage{gensymb}
\usepackage{float}
\usepackage{siunitx}

\usepackage{color}
\hypersetup{colorlinks,urlcolor=blue}
\usepackage{soul}

\usepackage{amsmath,amsthm,dsfont,amscd}

\DeclareMathOperator*{\argmax}{\arg\!\max}

\def\gsim{\mathrel{\raise.3ex\hbox{$>$\kern-.75em\lower1ex\hbox{$\sim$}}}}
\def\lsim{\mathrel{\raise.3ex\hbox{$<$\kern-.75em\lower1ex\hbox{$\sim$}}}}

\templatetype{pnasresearcharticle} 

\setboolean{displaywatermark}{false} 

\title{Intrinsically motivated collective motion}

\author[a]{Henry J. Charlesworth}
\author[a,b,2]{Matthew S. Turner} 

\affil[a]{Centre for Complexity Science, University of Warwick}
\affil[b]{Department of Physics, University of Warwick}

\leadauthor{Charlesworth} 

\significancestatement{Our study invokes a low-level principle that we believe might motivate animal behaviour in general and collective motion in particular - the principle that an agent seeks to {increase the number of} states that it is able to encounter in the future. This principle should confer evolutionary fitness for rather general reasons that we discuss. We report how the collective motion that emerges from this principle is similar to that seen in animal systems. In particular, coalignment, cohesion and collision avoidance all emerge naturally, even though none of these are encoded in the principle itself. In this sense, our work proposes a low level origin for the emergence of collective motion in animal systems.}

\authorcontributions{H.J.C. and M.S.T. designed research; H.J.C. performed research; H.J.C. and M.S.T. analyzed data; and H.J.C. and M.S.T. wrote the paper. }
\authordeclaration{The authors declare no conflict of interest.}
\correspondingauthor{\textsuperscript{2}To whom correspondence should be addressed. E-mail: m.s.turner@warwick.ac.uk}

\begin{abstract}
Collective motion is found in various animal systems, active suspensions and robotic or virtual agents. This is often understood using high level models that directly encode selected empirical features, such as co-alignment and cohesion. Can these features be shown to emerge from an underlying, low-level principle? We find that they emerge naturally under Future State Maximisation (FSM). Here agents perceive a visual representation of the world around them, such as might be recorded on a simple retina, and then move to maximise the number of different visual environments that they expect to be able to access in the future. Such a control principle may confer evolutionary fitness in an uncertain world by enabling agents to deal with a wide variety of future scenarios. The collective dynamics that spontaneously emerge under FSM resemble animal systems in several qualitative aspects, including cohesion, co-alignment and collision suppression, none of which are explicitly encoded in the model. A multi-layered neural network trained on simulated trajectories is shown to represent a heuristic mimicking FSM. Similar levels of reasoning would seem to be accessible under animal cognition, demonstrating a possible route to the emergence of collective motion in social animals directly from the control principle underlying FSM. Such models may also be good candidates for encoding into possible future realisations of artificial ``intelligent" matter, able to sense light, process information and move.
\end{abstract}

\dates{This manuscript was compiled on \today}
\doi{}

\begin{document}

\maketitle
\thispagestyle{firststyle}
\ifthenelse{\boolean{shortarticle}}{\ifthenelse{\boolean{singlecolumn}}{\abscontentformatted}{\abscontent}}{}

{
\dropcap{T}here have been notable recent advances in our understanding of collective motion motivated by thermodynamics or physical arguments
\cite{doi:10.1002/smll.200901976,Palacci936,Ramaswamy2017,0295-5075-101-2-20010,PhysRevLett.105.168103,RevModPhys.85.1143,PhysRevLett.75.4326,Bertin} and in animal systems \cite{cavagna,couzin_sample,fabry_penguins,PhysRevE.51.4282,Moussaid6884}. While generalised hydrodynamic theories \cite{RevModPhys.85.1143,PhysRevLett.75.4326,Bertin} can be obtained for certain active physical systems, the collective motion of agents capable of information processing can be far more complex.  For example, existing generalised hydrodynamic theories do not account for long-ranged interactions, such as are likely to arise in higher animals that rely on vision. Agent-based models have been developed that incorporate some of these potential complexities, e.g. distance-dependent attraction, orientation or repulsions \cite{PhysRevLett.105.168103,Ballerini1232, netherlands_bird_lady} or those relating more directly to vision \cite{hybridprojection,Gallup7245}. While these models have had some success in explaining animal data the starting point is usually an essentially empirical model. This leads to challenges, both in controlling against overfitting and providing low-level explanatory power, ``Why and how do agents co-align or remain in cohesive groups?". This question is difficult to answer if the model has co-alignment and cohesion hard-wired into it for essentially empirical reasons. 

We instead analyse an agent based system in which each agent senses, and then processes, information in the context of a predictive model of the future. It uses this model to determine its action in the present, re-computing its model of the future from scratch at each discrete time step. Each agent decides how to move according to a low-level motivational principle that we call {\it Future State Maximisation} (FSM): it seeks control in the sense that it maximises the variety of (visual) environments that an agent could access before some time horizon, $\tau$ time steps into the future. This is a form of control as it gives the agent many future options in a potentially uncertain world.

As we report below, FSM spontaneously generates collective motion of a sort that is similar to that observed in animal systems, i.e. moving, cohesive, highly aligned, swarms that are stable against small perturbations, see SI Movie~1. While there are even quantitative similarities with the structure and order in flocks of birds \cite{cavagna} the motivation for our work is not to mimic a particular animal system but rather to analyse a simple, low-level model that may provide a general conceptual basis for collective motion, here based on vision. Crucially, our model does not explicitly include co-alignment, cohesion or any other physical interaction, merely mutual visual perception between agents in infinite (2D) space.

There are several reasons why motivational principles like FSM, that loosely serve to {\it keep options open}, may confer fitness, either in artificial intelligence or in nature. FSM increases the control that an agent has over their future. Agents that have many options to re-position themselves relative to their neighbours, e.g. in response to the arrival of a predator, can likely better avoid or confuse that predator. 

In general, strategies like FSM that preserve an agent's freedom to reach many different outcomes in an uncertain world are expected to enhance fitness.

 Similar strategies are known to be successful in games like chess. Having access to many viable future lines of development is generically preferable, given uncertainty about how the game will actually develop. This confers robustness in defence and strategic manoeuvrability in attack. 
Chess players are familiar with the feeling of their options becoming progressively more limited as they lose a game, with the converse being strongly characteristic of winning. 
One attempt at formalising this kind of principle is the ``empowerment" framework which does so using the language of information theory \cite{empowerment2,capdepuy1,capdepuy2}. Our implementation  probably has most in common with this strand of the literature. FSM is an example of an {\it intrinsic motivation} \cite{empowermentintrinsic,intrinsicmotivationbook} where an incentive for behaviour is provided even in the absence of any specific tasks to complete or immediate  external rewards to be gained. Intrinsically motivated behaviour has been studied extensively in the psychology literature \cite{intrinsicmotivationwhite,intrinsicmotivations2} as well as more recently in the field of reinforcement learning \cite{intrinsicRL, empowermentRL, eysenbach2018} where it is used to aid exploration in environments where rewards are sparse. The key principle is that such behaviour should offer a generic and universal benefit to the agent, not because it is useful for solving any one particular problem, but because it is beneficial for a wide range of scenarios that the agent may encounter in the future.  A similar idea arises in the analysis of (hypothetical) causal entropic forces \cite{causalentropic,Hemminghaus}. These forces generate motion that increases an entropy-like measure of all paths into the future and can lead to behaviour with features usually thought to be characteristic of intelligence, including evidence for the spontaneous emergence of tool-use and social cooperation. 

Other work on decision making has some similarities with FSM \cite{Mann20150037}. In that study a formalism similar to \cite{causalentropic} was used to model agents making a group-level decision:  agents reach a consensus on a (single) decision; made in the same sensory context for all agents; without those agents perceiving states (more than a single step) in the future. No explicit dynamical model was defined or analysed in \cite{Mann20150037}. In the present work, FSM is applied to a group of agents that can move, perceive their own distinct environments and build independent models of future states that are accessible to them, guiding their decision making. This leads to the emergence of rich collective dynamics of a kind not previously realised.

Our work can also be seen as motivating the development of artificial particles that can sense, compute and move; so-called ``intelligent matter". This is a natural direction in which to develop existing active systems, e.g. phoretic colloids \cite{janus_phoresis}, swimming cells \cite{Kantsler1187} or active biological suspensions \cite{sanchez_nature} that have limited, and rigid, information processing capabilities. Having candidate algorithms to encode into this intelligent matter will help motivate its development. Heuristics that mimic FSM, as discussed below, may represent a particularly powerful choice for such algorithms. 

\section*{FSM applied to collective motion}
\section*{Methods}
We use deterministic computer simulation to study the motion of agents executing FSM. These agents are unit radius, phantom (i.e. able to overlap without repulsion) circular disks that are free to move on an infinite 2D plane. Their speed is the distance moved in each unit time step, with all lengths measured in disk radius units.
Fig~\ref{fig:movesandvisualstate}(a) shows the movement options available to each agent at each time step. These options are taken relative to its direction of motion in the previous time step. They are, in order: continue in the same direction with a choice of three different speeds, $v_0$ (nominal), $v_0-\Delta v$ (slow) or $v_0+\Delta v$ (fast). Alternatively, they are able to turn left or right by a small angle $\Delta\theta$, with speed $v_0$. Unless specified otherwise in what follows the nominal speed $v_0=10$, the speed variation $\Delta v = 2$ and the angular rotation $\Delta\theta=\text{\ang{15}}$.  At each time step the agent must choose one of these five actions $z \in \{z_1, z_2, z_3, z_4, z_5 \}$ and does so by executing a form of FSM, as described below.

\begin{figure}[!htb]
\centering
\includegraphics[width=1.0\linewidth]{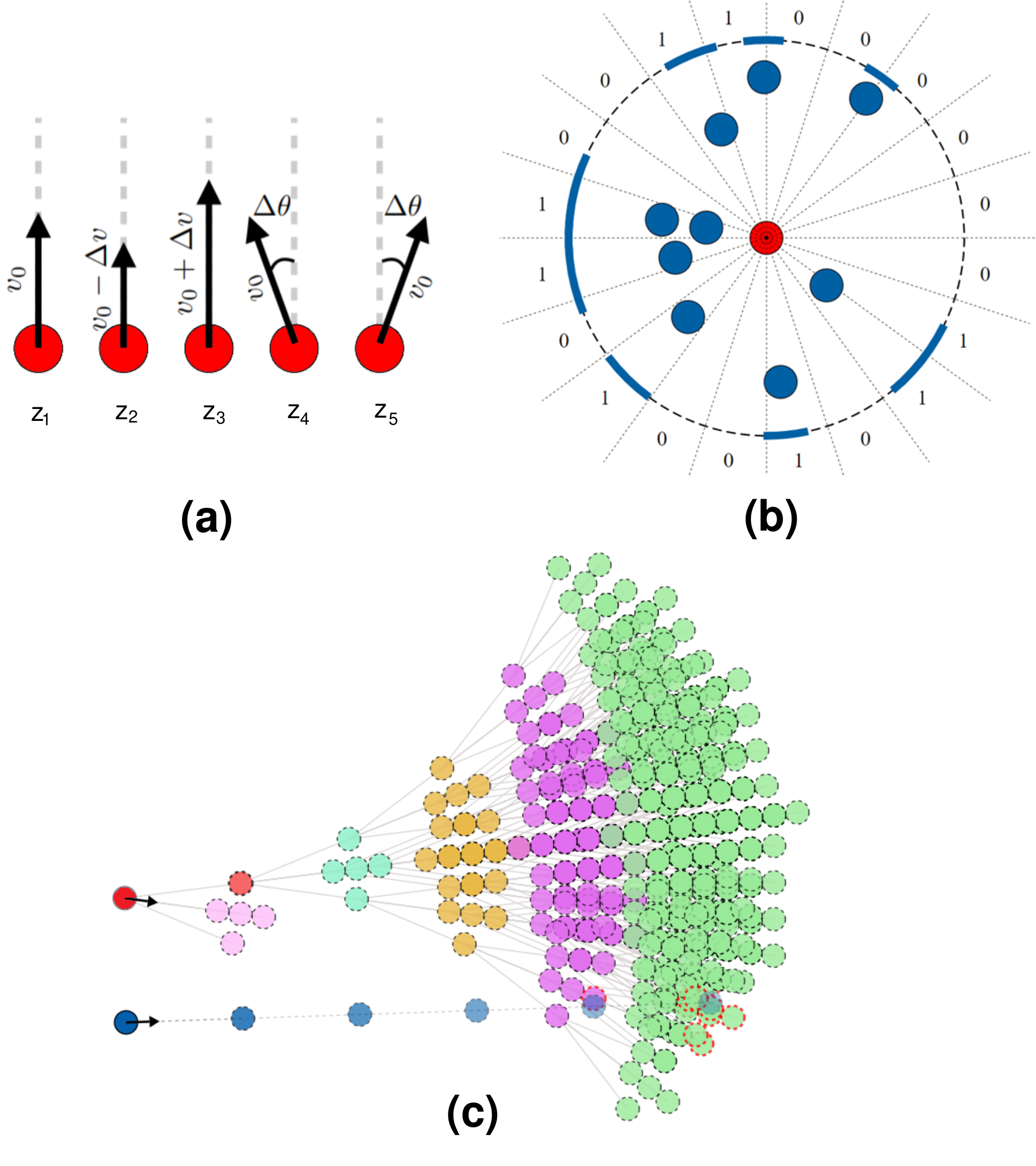}
\linespread{1.25}\selectfont
\caption{\small Sketch showing an agent's movement options, a representation of the visual state of the world around it and its future decision-tree. (a) The five actions available to each agent at every time step, given that its previous move was in the direction of the dashed line, continue in the same direction at a nominal/slow/fast speed, or turn left/right, respectively. (b) A representative agent (red) sees the other agents (blue) geometrically projected onto a retina-like sensor array. Each sensor registers 1 if a line of sight through more than half of its angular region intersects other disk(s), corresponding to the solid blue regions on the perimeter; 0 otherwise. 
This $n_s$-dimensional Boolean vector is the agent's sensory input and represents its ``state". Here we show $n_s=20$, for clarity. (c) The spatial positions that an agent, shown as red, can access in the future form nodes on a fan-like tree, colour coded by the time into the future: pink/red (1 step), cyan (2 steps), orange (3 steps), magenta (4 steps), green (5 steps); in this cartoon the maximum future time horizon is therefore $\tau=5$. The branch of this tree that the agent explores is contingent on its next move (here shown as a turn to the left, in red). A similar branch exists for the four other possible moves but these are omitted, for clarity. The red agent computes the future sensory states accessible to it at each future node, as described in (b), choosing the move in the next time step that leads to the branch with the largest {\bf number} of {\bf distinct} visual states. The nodes that are highlighted in dotted red correspond to positions that the agent anticipates will overlap (``collide") with other agents. Here a single other colliding agent is shown in blue, for clarity. When computing the number of distinct visual states we exclude those from nodes that correspond to, or follow after, such a collision.}
\label{fig:movesandvisualstate}
\end{figure}

Fig~\ref{fig:movesandvisualstate}(b) shows how each visual state is constructed using the positions of the agents. 

This visual state is constructed for each agent by geometrically projecting all $N-1$ other disks down onto its centre. This involves constructing pairs of lines that each pass through the focal disk's centre and are tangent to both sides of the other disks. Each of these lines can be specified by an angle, measured relative to the agent's direction of motion. This allows us to define angular regions in which a line of sight will intersect with one or more other disks, shown as solid blue regions on the perimeter of
Fig~\ref{fig:movesandvisualstate}(b). We construct $n_s$ discrete visual sensors that each relate to an angular region of size $2\pi/n_s$. The radial dotted lines in Fig~\ref{fig:movesandvisualstate}(b) denote the angular sensors (not the tangent lines). Each sensor registers 1 if more than half occupied by angles along which a line of sight will intersect other disk(s), i.e. the fraction of solid blue; 0 otherwise. Unless stated otherwise $n_s=40$ in all simulations. Panel (c) shows how each agent constructs its future decision tree, {\it given a model for the motion of all other agents}, here simply that they continue on their previous trajectory at nominal speed $v_o$, as illustrated by the blue agent. In this way the agent can compare each of the five moves available to it based on the {\bf absolute number} of {\bf different} visual states on all nodes accessible from that move. {\bf It chooses the move that maximises this measure}. 

In more mathematical language, we define the visual state $\mathbf{f_i} \in \{0, 1\}^{n_s}$ of an agent on the $i^{\rm th}$ node of its tree of potential future states, as discussed above. Each of the five available moves in the next time step leads to branch $\alpha$ of the tree of potential future states. For each of these five branches we then construct a set $S_{\alpha}$ consisting of all of the \textit{unique} visual states $\mathbf{f_i^{(\alpha)}}$ within that branch. The future time horizon (tree depth) is $\tau=4$ in our simulations, unless stated otherwise. Each branch is then given a weight $W_{\alpha} = | S_{\alpha} |$, and the agent then chooses the current action  $z_{\\alpha^*}$ such that $\alpha^* = \argmax_{\alpha} z_{\alpha}$.

Consider a toy example of this process in which there are only $n_s=4$ sensors and two possible actions. Imagine that the branch $\alpha=1$, following action $z_1$, leads to three nodes with visual states of $\{1,0,1,0\}$, $\{1,0,0,0\}$, $\{1,0,0,0\}$ and $\{1,0,1,0\}$, while the branch $\alpha=2$, following action $z_2$, leads to four nodes with visual states of $\{1,0,1,0\}$, $\{1,0,0,0\}$, $\{1,1,0,0\}$ and $\{1,0,1,1\}$. In this example, branch $\alpha=2$, and hence action $z_2$, would be chosen because it leads to a future with four {\bf distinct} Boolean vectors (states) whereas the branch $\alpha=1$ contains only 2 distinct states; the vectors $\{1,0,1,0\}$ and $\{1,0,0,0\}$ being repeated.

Some nodes on the decision tree correspond to collisions and are highlighted on Fig~\ref{fig:movesandvisualstate}(b) with a dotted red outline. An agent considers any branch of its decision tree to terminate on collision, i.e. this and any subsequent nodes are deemed inaccessible. In this way the agent tends to avoid collisions because they contribute no states to its future. We find a strong reduction of collisions in the FSM trajectories that result, typically 2-3 orders of magnitude below a control collision rate (see Fig 2 in the SI). This is in spite of the fact that there is no explicit suppression of collisions, e.g. via physical interactions.

In the SI we discuss how to generalise this to a continuous measure of the degeneracy of future visual states. 

In \cite{causalentropic} a Gibbs measure of the accessible state space, rather than a count the number of distinct states, is used to quantify the future freedom. Our work could be extended in this direction in the future.

\begin{figure}[!htb]
\centering
\includegraphics[width=1.0\linewidth]{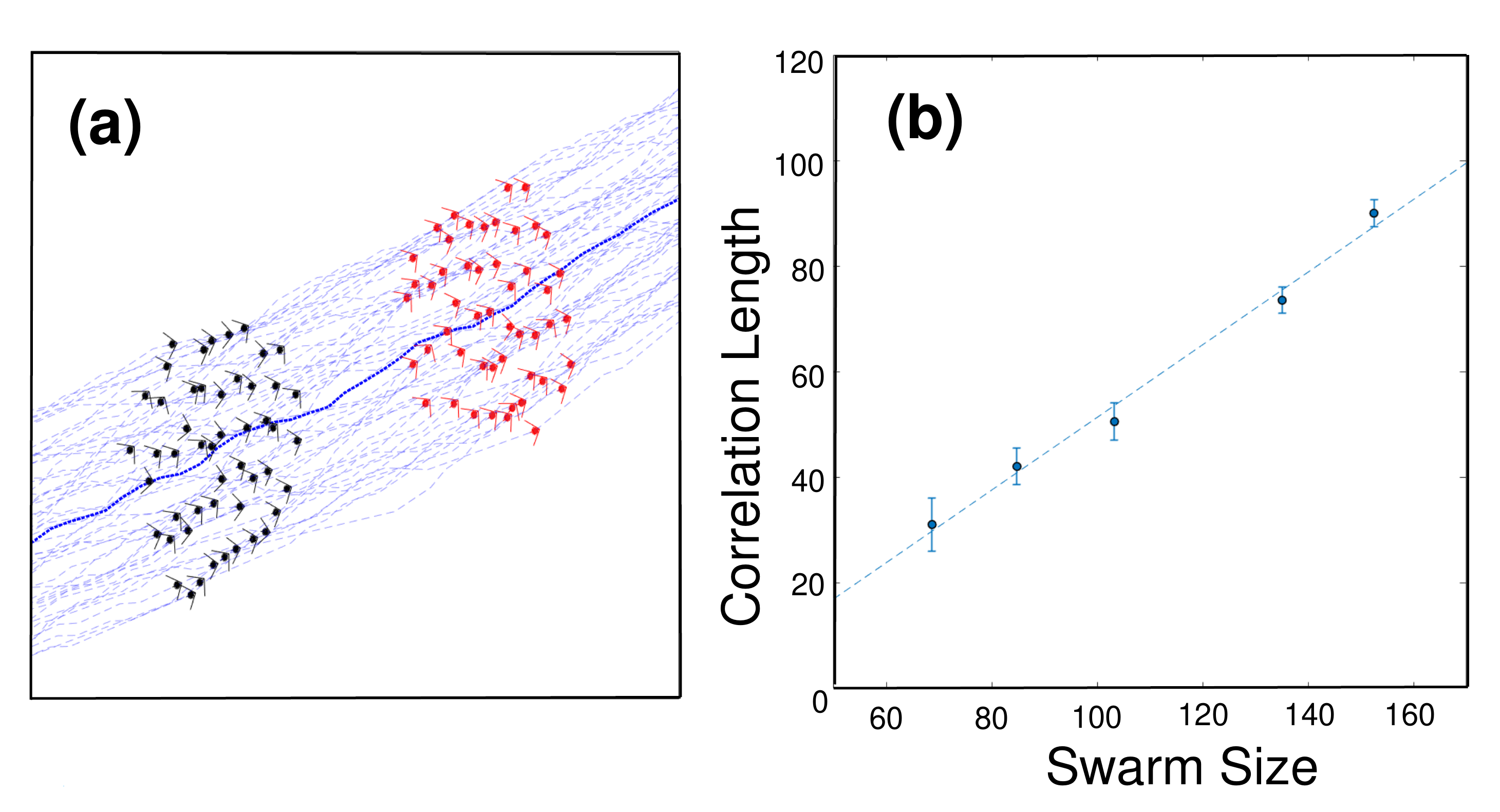}
\linespread{1.25}\selectfont
\caption{\small (a) Structure of collective swarms that emerge under FSM dynamics, as described in Fig 1(a). Snapshots of a typical realisation at two different times showing the trajectories of the agents (light dashed lines) and centre of mass (dark dotted line), with $N=50$, $n_s=40$, $v_0=10$, $\Delta v=2$, $\Delta\theta=15^\circ$ and a time-horizon of $\tau=4$. Wedges show agents' direction of motion, see SI Movie 1. (b) The centre-of-mass frame velocity correlation function for agents is computed for systems with the same parameter values except that the data points correspond to $N=50, 75, 100, 150, 200$ agents. Shown is the correlation length thereby obtained, here defined as the distance at which this correlation function crosses zero (nearby agents are positively correlated, distant ones are negatively correlated). This correlation length is compared against the corresponding swarm size, the square root of the area of a convex hull containing all agents. See SI for details.}
\label{fig:snap}
\end{figure}

\begin{figure*}[!htb]
\centering
\includegraphics[scale=0.7]{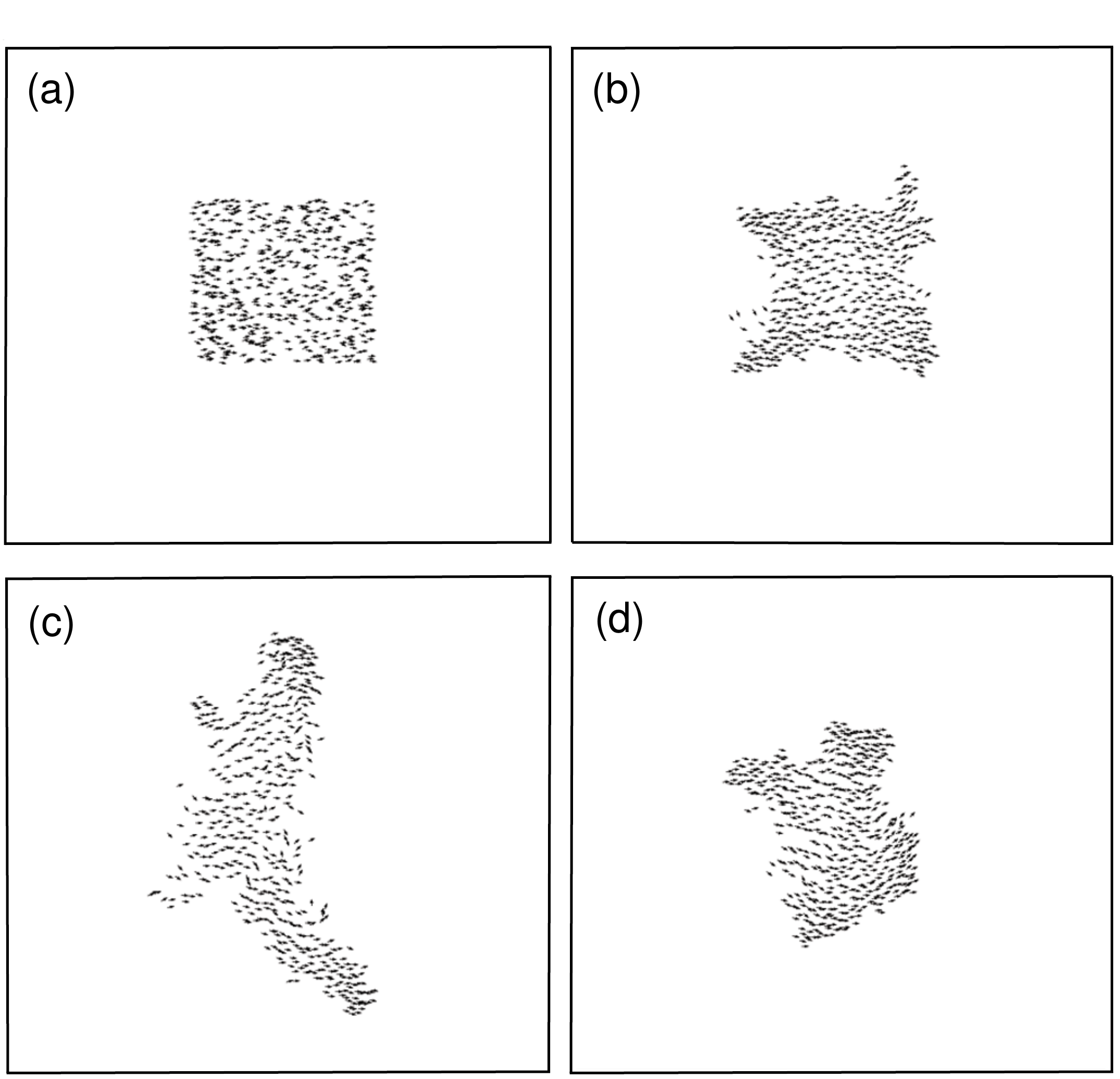}
\linespread{1.25}\selectfont
\caption{\small Snapshots of a swarm made up of $N=500$ agents with $\tau=5$, shown at different times 
in a frame co-moving with the swarm's centre of mass. Panel (a) shows the initial state of the swarm and then (b), (c) and (d) show snapshots of its subsequent evolution (in chronological order). In this example we use a continuous measure of visual degeneracy (see SI for details). The full simulation is shown in SI Movie~4.}
\label{fig:N500snapshots}
\end{figure*}

\section*{Results}
Swarms similar to that shown in Fig~\ref{fig:snap}(a) arise from these FSM dynamics across a broad range of parameter values, see SI for a comparison. However, there are some restrictions, e.g. the number of sensors can neither be too large (so that all states become unique) or too small (sensory resolution is lost) and the time horizon must be sufficiently long. For time horizons that are too short ($\tau<4$ for $N=50$) the swarm becomes less stable with agents separating from the main swarm increasingly frequently. In general the initial conditions must be chosen to be roughly commensurate with the steady state. If the system is prepared in an initial configuration from which the agents' decision trees cannot perceive the steady state within $\tau$ time steps then the swarm fragments, typically into cohesive subgroups, see SI Movie 2 for an example of this phenomenon with $N=500$.  Such initial conditions correspond to widely separated and/or orientationally disordered agents. Robustness to variation of the initial conditions improves with increasing $\tau$.

The state shown in Fig~\ref{fig:snap}(a)  and in SI Movie 1 has further qualitative similarities with animal systems and, in particular, large flocks of starlings: Its alignment order parameter is within 1\% of a typical value for starling flocks \cite {cavagna} and it is in a state of {\it marginal opacity}, in which the fraction of sensors in state 0 to state 1 is order unity \cite{hybridprojection} (see figure 2 in the SI for more details). Finally, the correlation length scales with the system size, as shown in panel (b). This is indicative of scale-free correlations, another feature of starling flocks \cite {cavagna}, and systems close to criticality more generally\cite{Bialek_criticality}. Fig~{\ref{fig:snap}}(c) shows snapshots of a larger swarm ($N=500$, $\tau=5$), sequentially in time,  with motion determined by FSM on the continuous measure of visual state degeneracy described in the SI. Whilst for smaller swarms the two approaches give virtually identical results (compare SI movies 1 and 3), for larger swarms the continuous measure has more variety in its steady state collective dynamics and is more robust to fragmentation (contrast SI movies 2 and 4).

It is perhaps somewhat counter-intuitive that such a highly ordered state emerges, given that FSM can be interpreted as preferring highly varied (roughly, high entropy) distributions of states. This is because FSM is insensitive to the variety/disorder of the swarm {\it in the present}. It is from such a highly ordered state that the swarm can access the greatest variety of states {\it in the future}. Thus it targets this state and remains there. The state is cohesive because nearby agents then have the most freedom to re-arrange their relative positions, Marginal opacity is selected because most configurations have sensor states roughly evenly split between 0 and 1.

\begin{figure}[!htb]
\centering
\includegraphics[width=1.0\linewidth]{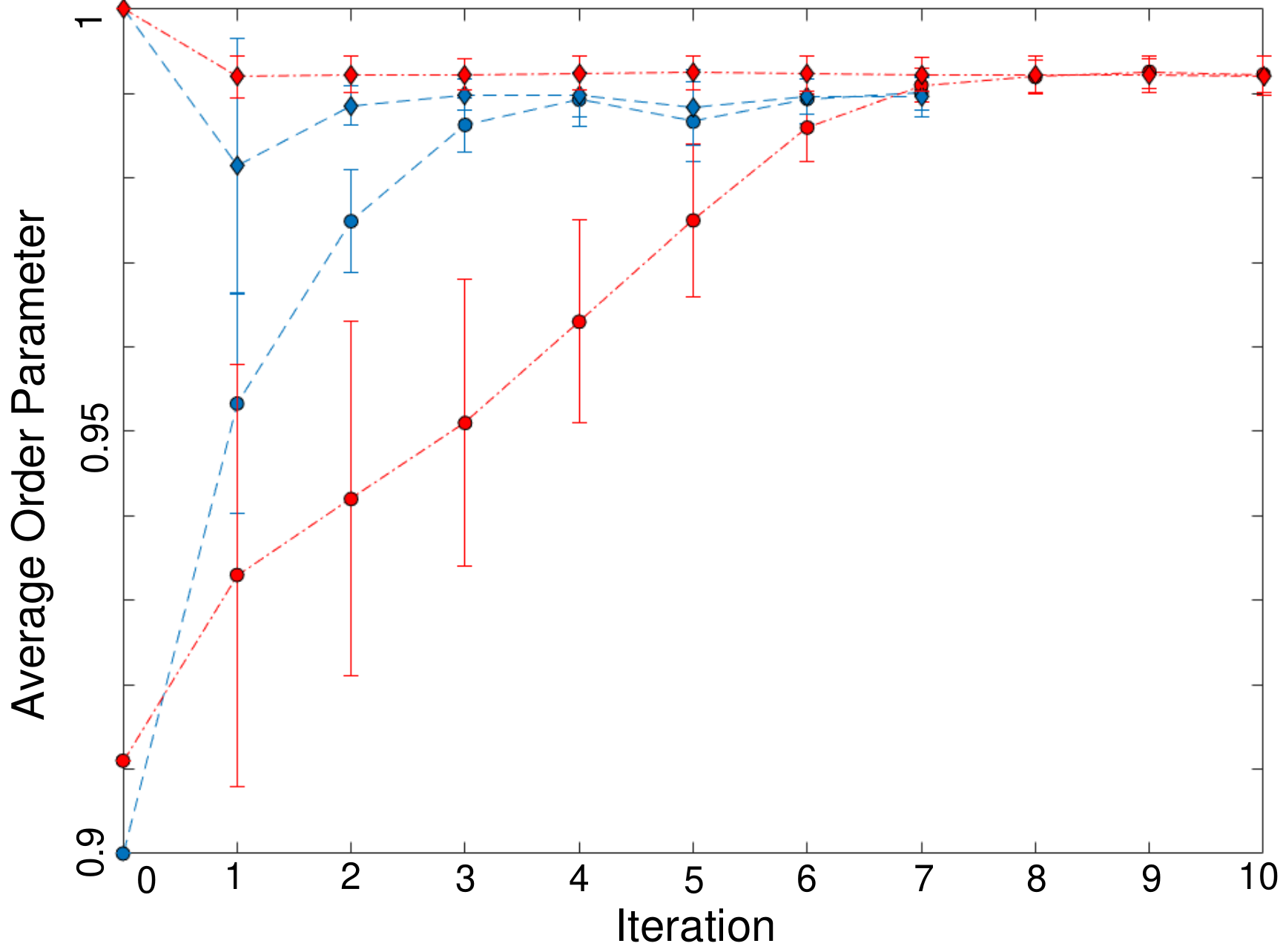}
\linespread{1.25}\selectfont
\caption{\small Convergence of heuristic A (order targeting, blue) and B (topological Vicsek \cite{PhysRevLett.105.168103}, red) to a value of the order parameter that is self-consistent with the value realized by FSM in each case. An initial (iteration 0) order parameter for the heuristic ($\phi_A$ and $\phi_{B}$, respectively) is chosen. This parameterises the model of all (other) agents to be used when constructing their trajectories into the future in order to apply FSM on each agent's predicted future visual states. The average order realised by the FSM simulation then serves as the order parameter for the heuristic in the next iteration and the process is repeated. The order converges, both from above and below, to an average order parameter that is the same, both for the heuristic and the motion generated by FSM using that heuristic to model the behaviour of other agents. FSM under heuristic A is unstable for values of $\phi_A\lsim0.9$, leading to flock fragmentation into (ordered) subgroups. 
Parameter values as given in Fig~\ref{fig:snap}. See also SI Movie 5.}
\label{fig:convergence}
\end{figure}

\section*{Changing the heuristic used to model hypothetical future trajectories}
A key ingredient of the FSM model is an assumption for how the other agents will move in the future. Without such an assumption their future positions remain undetermined and the corresponding visual projections cannot therefore be computed. Fig~1(c) shows the simplest of four different assumptions, or {\it heuristics}, that we report in this letter: all other agents (only a single (blue) one is shown) are assumed to continue on {\bf ballistic} trajectories, without turning, at speed $v_o$. The structure of the cohesive, co-aligned swarms that spontaneously emerge under this assumption are shown in Fig~\ref{fig:snap}(a) (see also SI Movie 1). The ballistic motion assumption is an approximate model for the motion of the other agents  and is not strictly self-consistent insofar as all agents are identical and actually move according to FSM. Hence, the (other) agents won't move in exactly such a ballistic fashion, as can be seen from the individual trajectories in Fig~\ref{fig:snap}(a). Nonetheless, this assumption is quite good for the highly ordered (co-aligned) swarms that do emerge from FSM. All agents would indeed continue moving in exactly the same direction under perfect co-alignment. The alignment order parameter is here defined as $\phi=\langle \frac1N\sum_i^N \hat {\bf v}_i(t)\rangle$ with the average performed over  time and $\hat {\bf v}_i(t)$ a unit vector in the direction of motion of the $i^{\rm th}$ agent at time step $t$. The swarm in Fig~\ref{fig:snap}(a) has order $\phi \simeq 0.98$.

Other heuristics can be made self-consistent with FSM. Examples include: (A) Agents are assumed to collectively target a particular value of order. At each time step every agent, in random order, turns in either direction if this brings the collective order closer to the target order $\phi_A$, otherwise continuing at speed $v_o$. (B) Agents are assumed to move at speed $v_o$ according to a topological version of the Vicsek model \cite{PhysRevLett.105.168103}, in which co-aligning neighbours are those that share edges under a Delaunay triangulation. As usual this model involves a variable noise $\eta$, with a one-to-one relationship between this and the average order parameter at that noise $\phi_{B}(\eta)$. Fig~\ref{fig:convergence} shows that both these heuristics can be made self-consistent with FSM at the level of the order realized: 
The FSM trajectories that are generated, using these heuristics as a model for the motion of all (other) agents, then have the same order as is generated by the bare heuristic, a value that was not known {\it a priori}. Any evolutionary pressure to adopt FSM-like dynamics should, presumably, also favour the ability to self-consistently predict the behaviour of other members of the group in this way.

\section*{Training a neural network to mimic the FSM algorithm}
While the full FSM algorithm is computationally demanding, an artificial neural network could serve as an example of a heuristic that can closely mimic FSM, and fitness benefits arising therefrom. Crucially, once trained, it is computationally simple and fast. Similar levels of reasoning could be expected to operate under animal cognition. We do not claim that an artificial neural network would be a direct model for (wet) neural networks, even though the former field was indeed motivated by the latter. We only argue that reasoning with this level of complexity could be encoded in an animal brain. This heuristic, like the others described above, could also be used as a model for the behaviour of other agents during FSM. 

We have trained a multi-layered neural network on the simulated trajectories that arise under FSM over 200,000 time steps, as sketched in Fig~\ref{fig:nnstructure}. \begin{figure*}[!htb]
\centering
\includegraphics[scale=0.5]{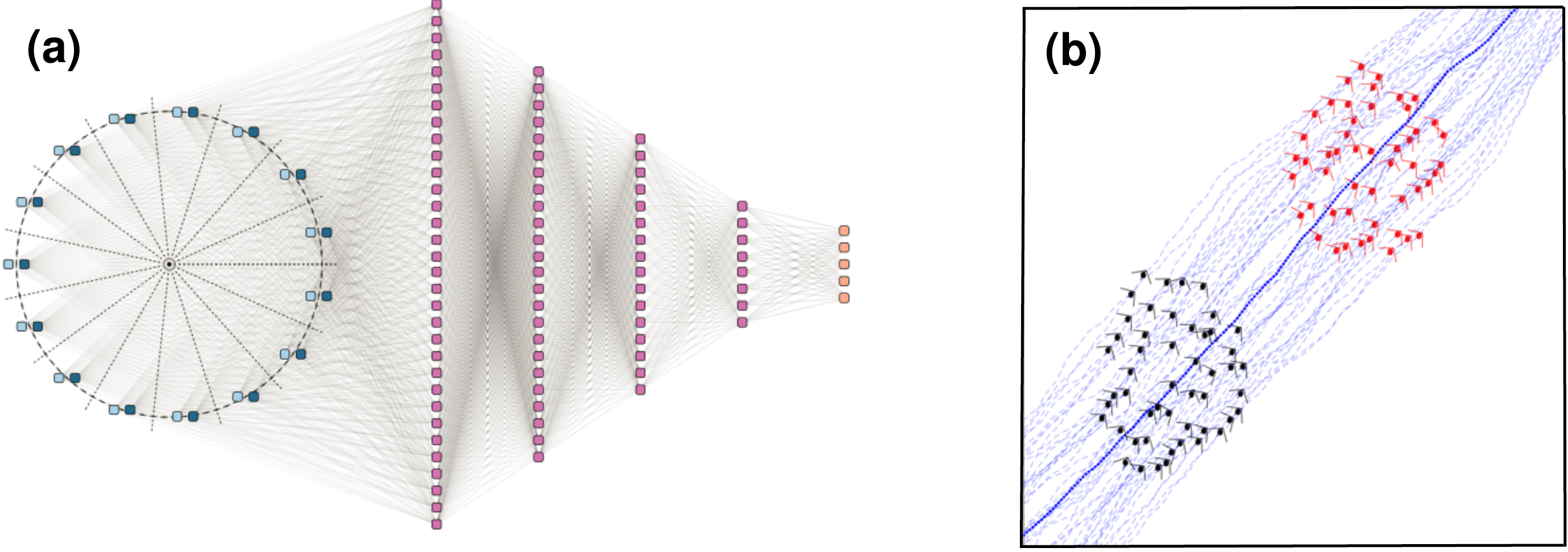}
\linespread{1.25}\selectfont
\caption{\small Training a neural network as a heuristic approximating FSM. (a) Sketch of the network architecture. The network takes as its input the agent's current speed and the state of each sensor in both the current and previous time steps, represented as light and dark blue squares on each sensor (left). This is then passed through four hidden layers of neurons of sizes 200, 100, 50 and 25 which have {\tt RelU} activation functions. These are  attached to a {\tt softmax} classifier which outputs an integer between 1 and 5, identifying the next move (final output, right). The network was trained to mimic FSM trajectories using 10 million examples (data from 200,000 simulation time steps). (b) The output dynamics from this network is seen to closely mimic the FSM trajectories shown in Fig~\ref{fig:snap}, see SI Movie 6.}
\label{fig:nnstructure}
\end{figure*}
We gather training data by running the full FSM algorithm using nominal parameters ($N=50$, $\tau=4$, $n_s=40$, $\Delta\theta = \text{\ang{15}}$, $v_0=10$, $\Delta v = 2$). Future visual states are computed under the assumption that other agents will move ballistically in their future trajectories, i.e. at speed $v_0$ in their current direction of motion. We generate 800 separate simulations, each with agents initially placed randomly in a square region with dimensions that vary between 80 and 160 disk radii. Each agent's initial orientation is chosen randomly from a Gaussian distribution with mean orientation along the nominal $x$-direction and a standard deviation of $2 \Delta \theta$. We chose these different initial conditions to provide representative examples of trajectories that recover from perturbations. This allows the trained network to make decisions that mimic FSM in situations that vary from the steady state, improving its robustness. In each of the simulations we record the current speed and the current \textit{and previous} visual state of every agent at every time step, along with the actual decision made by the FSM algorithm in that situation (represented as an integer between 1 and 5). Note that including memory, via the previous visual state, is found to be crucial in order to train a network which qualitatively reproduces the behaviour of the full FSM algorithm. The training process is a supervised learning problem in which we have 10 million labelled example decisions each corresponding to a vector input of dimension 81 (2 $n_s + 1$, for the speed) with each output an integer between 1 and 5.

The neural network architecture we employ consists of a hidden layer of 200 fully connected neurons connected to the input with three further fully connected hidden layers of sizes 100, 50 and 25 respectively. The last of these is connected to a softmax classifier which outputs an integer between 1 and 5 corresponding to the selected action. All of the hidden layers use the RelU activaton function. We trained the neural network on all of the data for 500 epochs using the ADAM optimizer under Keras with an initial learning rate of 0.0001.

The output from our artificial neural network is seen to closely mimic the FSM trajectories, see SI Movie 6. 

In summary, we propose a form of intrinsically motivated collective motion based on Future State Maximization (FSM). This involves a minimal representation of vision in which agents seek to increase their control of the visual world around them. Specifically, they target being able to reach the greatest variety of future environments. The potential fitness benefits of this lie in the fact that it gives the agent freedom to access different outcomes in an uncertain world. Cohesive, ordered swarms that resemble natural animal systems spontaneously emerge under FSM. This behaviour can be encoded in heuristics, mimicking full FSM. A neural network is an example of the kind of heuristic that could mimic FSM under animal cognition, providing a possible route for the evolutionary selection of this behaviour. Such heuristics could also lie within the processing power of future realisations of ``intelligent" materials that may incorporate sensors, as well as the ability to move.

\acknow{This work was partially supported by the UK EPSRC though the Mathematics for Real-World Systems CDT, EP/L015374/1 (H.J.C.). Computing facilities were provided by the Scientific Computing Research Technology Platform of the University of Warwick. We acknowledge stimulating discussions with George Rowlands (Warwick) and Hugues Chat\' e (Saclay).}

\showacknow{} 

\bibliography{references}

\end{document}